\documentstyle[11pt,epsfig]{article}
\title{\bf Solutions of N-dimensional Schr\"{o}dinger Equation with Morse Potential Via Laplace Transforms}
\author{S. Miraboutalebi$^{1}$\thanks{Tel.: +98 021 77317701-9;
Fax: +98 021 77317716; E-mail
address:~smirabotalebi@gmail.com} \,  and  L. Rajaei$^{2}$\\
{$^1$\small Department of Physics, Islamic Azad University, North
Tehran Branch, Tehran, 1651153311, Iran.}\\$^2${\small Physics
Department, Qom
University, Qom, Iran}\\
}
\begin{document}
\maketitle
\begin{abstract}

A study is undertaken to investigate an analytical solution for
the N-dimensional Schr\"{o}dinger equation with the Morse
potential based on the Laplace transformation method. The results
show that in the Pekeris approximation, the radial part of the
Schr\"{o}dinger equation reduces to the corresponding equation in
one dimension. Hence its exact solutions can be obtained by the
Laplace transformation method of G. Chen, phys. Lett. A 326 (2004)
55. In addition, a comparison is made between the energy spectrum
resulted from this method and the spectra that are obtained from
the two-point quasi-rational approximation method and the
Nikiforov-Uvarov approach.

\end{abstract}


%%%%%%%%%%%%%%%%%%%%%%%%%%%%%%%%%%%%%%%%%%%%%%%%%%%%%%%%%%%%%%%%%%%%%%%%%%%%%%%

\section{Introduction}
The vibration and rotational movements of the diatomic molecules
are topics of research in a wide range of scientific fields
including molecular physics and astrophysics \cite{CY}. However,
these areas are considered to be the main tools for other
disciplines such as biology and environmental science \cite{HW,
FM}. One of the most available methods to describe the vibration
of the diatomic and even polyatomic molecules is the Morse
potential \cite{Morse}. The rotation of molecules presents
independently with the centrifugal potential and involves the
quantum number of the angular momentum, namely $\ell$.

Solutions of the Schr\"{o}dinger equation for the sate $\ell=0$
has been found in \cite{Morse}-\cite{IS}. However, due to
complexity of the case $\ell\neq 0$, the wave equation can only be
solved by using perturbation and approximation.  The most employed
estimation to solve the equation is the Pekeris method,
\cite{Pe,Flugge}. This method is suitable for obtaining the local
solutions of the equation, when the range of the nuclear distance
is not far from its equilibrium position. Moreover, this method
induces some restriction on the upper limit of the quantum number
$\ell$.

A number of methods have been proposed to solve the
Schr\"{o}dinger equation with the Morse potential for $\ell\neq 0$
case, among which are the factorization scheme \cite{IH}-\cite{
DLF}, the path integral formulation \cite{G1}-\cite{KS}, the super
symmetry approach \cite{CKS}-\cite{Gran}, the algebraic way
\cite{BGQ}-\cite{GDD}, the power series expansion
\cite{JDS}-\cite{Z}, the two-point quasi-rational approximation
method \cite{CPM, CMP}, the $1/N$ expansion procedure
\cite{Lai}-\cite{Bag}, the transfer matrix method
\cite{OCS}-\cite{NA}, the asymptotic iteration method
\cite{CHS1}-\cite{BB} and Nikiforov-Uvarov approach \cite{BH, SI}.

One of the most effective methods for solving the Schr\"{o}dinger
equation with different sort of spherically symmetric potentials
is the Laplace transformation method \cite{Kre}. The advantage of
this method is that a second order differential equation reduces
to a first order differential equation. It was Schr\"{o}dinger who
used this technique for the first time in quantum physics in order
to solve the radial eigenfunction of hydrogen atom,  \cite{Sch}.
The method has become commonly employed ever since to solve
various kind  of the spherically symmetric potentials \cite{En1}-
\cite{PC}.

The Laplace transform method also has been applied to solve the
one dimensional Schr\"{o}dinger equation with the Morse potential,
when $\ell=0$, by Chen in \cite{Chen1}. In the proposed procedure
the exact bound state solutions are obtained in an effective
manner. The present paper attempts to solve the radial
Schr\"{o}dinger equation for the Morse potential in the Pekeris
approximation with the Laplace transform method, in three and then
in $N$ dimensions.

This paper is organized as follows: In  section  two the Pekeris
approximation is reviewed. In section three the bound state
solutions in three dimensions are obtained. In this section it is
shown that the three-dimensional Schr\"{o}dinger equation with the
Morse potential can be reduced to the corresponding equation in
one dimension. As a result, its exact solutions can be achieved by
the Laplace transformation method of \cite{Chen1}. Furthermore, a
comparison is made between the resulted energy spectrum and those
spectra that are obtained by using the two-point quasi-rational
approximation method and the Nikiforov-Uvarov approach. In section
four the procedure is generalized to an arbitrary dimension $N$
and the bound state solutions of the N-dimensional Schr\"{o}dinger
equation for the Morse potential are found. Finally, section five
presents the results.

%%%%%%%%%%%%%%%%%%%%%%%%%%%%%%%%%%%%%%%%%%%%%%%%%%%%%%%%%%%%%%%%%%%%%%%
\section{Schr\"{o}dinger equation for Morse potential with rotation correction}

The time independent Schr\"{o}dinger equation for an arbitrary
potential $V(\textbf{\textsf{r}})$ is given by

\begin{equation}\label{e1}
\frac{-\hbar^2}{2m}
\nabla^{2}\psi(\textbf{\textsf{r}})=[E-V(\textbf{\textsf{r}})]\psi(\textbf{\textsf{r}})\,.
\end{equation}
For spherical symmetric potential, the wave function
$\psi(\textbf{\textsf{r}})$ can be separated as \cite{JDS}

\begin{equation}\label{e2}
\psi(\textbf{\textsf{r}})=\frac{1}{r}
R_{\ell}(\textsf{r})Y^{\ell}_m(\theta,\phi)\,.
\end{equation}
Substituting Eq.(\ref{e2}) in Eq.(\ref{e1}), the equation for the
radial wave function becomes

\begin{equation}\label{e3}
\frac{-\hbar^2}{2m}\left[\frac{d^2}{d
\textsf{r}^2}-\frac{\ell(\ell+1)}{\textsf{r}^2}\right]R_{\ell}(\textsf{r})=[E_{\ell}-V(\textsf{r})]R_{\ell}(\textsf{r})\,.
\end{equation}
The Morse potential \cite{Morse, Flugge}, has the following form:

\begin{equation}\label{e4}
V(\textsf{r})=D\left(e^{-2\alpha r}-2e^{-\alpha r}\right)\,,
\end{equation}
where

\begin{equation}\label{e5}
r=\frac{\textsf{r}-\textsf{r}_{0}}{\textsf{r}_{0}}\,,
\end{equation}
and $\textsf{r}_{0}$ is the equilibrium position of molecules. The
parameter $D$ describes the depth of the potential and the
dimensionless parameter $\alpha$ characterizes the potential
acting range.

From the classical point of view, the nuclear distance
$\textsf{r}$ even for high fluctuation levels, will not oscillate
significantly far from the equilibrium distance $\textsf{r}_{0}$.
Hence it is reasonable that
$r=|\frac{\textsf{r}-\textsf{r}_{0}}{\textsf{r}_{0}}|\ll 1$. This
allows to expand $\frac{1}{(1+r)^2}$ of the centrifugal potential
term in Eq.(\ref{e4}), in the form

\begin{equation}\label{e6}
\frac{1}{(1+r)^2}=1-2r+3r^2-4r^3+...\,,
\end{equation}
and considering its first few terms.  Up to order $r^3$, this
expansion can be replaced by, \cite{Flugge}
\begin{equation}\label{e7}
\frac{1}{(1+r)^2}\cong C_{0}+C_{1}e^{-\alpha r}+C_{2}e^{-2\alpha
r}\,,
\end{equation}
since the power series expansion of the later relation yields the
former, with the following definitions

\begin{equation}\label{e8}
 C_{0}=1-3/\alpha+3/\alpha^2,\,\, C_{1}=4/\alpha-6/\alpha^2,\,\,
 C_{2}=-1/\alpha+3/\alpha^2\,.
\end{equation}
Substituting Eq.(\ref{e7}) into Eq.(\ref{e4}), the Schr\"{o}dinger
equation (\ref{e4}) becomes

\begin{equation}\label{e9}
\left(\frac{d^2}{d r^2}- \eta^2 e^{-2\alpha r} +  2\zeta^2
e^{-\alpha r}- \beta_{1}^2\right) R_{\ell}(r)=0\,,
\end{equation}
where

\begin{equation}\label{e10}
\begin{tabular}{c}
$\beta_{1}^2 = -\frac{2m
\textsf{r}_{0}^2}{\hbar^2}E_{\ell}+\ell(\ell+1)C_{0}$\,,
\\\\
$\zeta^2  = \frac{4m
\textsf{r}_{0}^2}{\hbar^2}D-\frac{\ell(\ell+1)}{2}C_{1}$\,,
\\\\
$\eta^2  = \frac{2m
\textsf{r}_{0}^2}{\hbar^2}D+\ell(\ell+1)C_{2}$\,.
\end{tabular}
\end{equation}
Eq.(\ref{e9}) is the three-dimensional Schr\"{o}dinger equation
for the Morse potential in the Pekeris approximation.

It is noticeable that the validity of the approximation (\ref{e7})
depends on the order of magnitude of the relative distance $r$ and
likewise by considering Eq.(\ref{e8}), the strength of the
parameter $\alpha$. In order to justify the fitness of the L.H.S
and the R.H.S of Eq.(\ref{e7}) for different magnitude of $r$ and
strength of $\alpha$, these terms are plotted in Fig. (1) where
the solid line corresponds to the centrifugal term $1/(1+r)^2$
without the factor $\ell(\ell+1)$, the other lines show $(C_{0}+
C_{1}e^{-\alpha r}+ C_{2}e^{-2\alpha r})$ for different strength
of $\alpha$. According to the figure, by increasing of $\alpha$
the approximation ($\ref{e7}$) would be correct in smaller domain
of $r$. In the worst case, namely for $I_{2}$, at $r=0.15$ the
relative discrepancy is only about $0.0136$. Please note that the
validity of the Pekeris approximation also depends on the
magnitude of the rotational quantum number $\ell$. In fact the
relative discrepancies are multiplied by the factor
$\ell(\ell+1)$. Therefore it can be concluded that the Pekeris
approximation is not reliable for higher values of $\ell$.

\begin{figure}
\centerline{\epsfig{figure=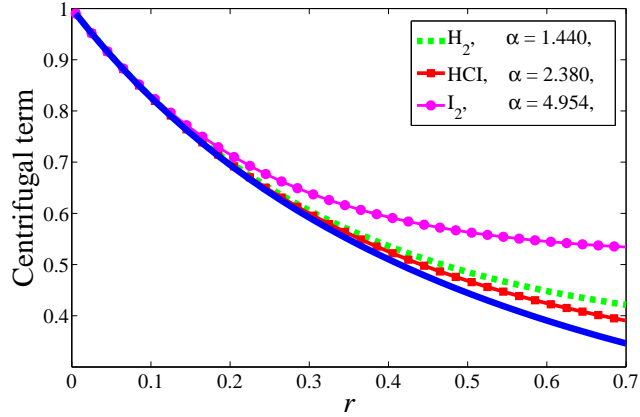,width=9.5cm}} \caption{{\small
The centrifugal term for different values of $\alpha$ versus the
relative distance r.}}
\end{figure}

%%%%%%%%%%%%%%%%%%%%%%%%%%%%%%%%%%%%%%%%%%%%%%%%%%%%%%%%%%%%%%%%%%%
\section{Bound state solutions in three dimensions}

Considering only the fluctuation modes, that is the case $\ell=0$,
the radial Schr\"{o}dinger equation has been solved in
\cite{Chen1} via Laplace transforms. The more general case
$\ell\neq 0$ contains the rotational energies in addition to the
vibrational modes, and the corresponding radial Schr\"{o}dinger
equation reduces to Eq.(\ref{e9}). This section shows that by
following the procedure of \cite{Chen1}, the eigenvalues and
eigenfunctions of this case can be obtained via Laplace
transforms.

First, a new variable is defined as
\begin{equation}\label{l1}
y= \frac{2\eta}{\alpha} e^{-\alpha r}\,.
\end{equation}
By applying this new variable in Eq.(\ref{e9}), the following
equation results

\begin{equation}\label{l2}
\left(y^2\frac{d^2}{d y^2}+y\frac{d}{dy}-\frac{1}{4}y^2 +
\frac{\kappa}{2}y -\beta^2\right)R_{\ell}(y)=0\,,
\end{equation}
where

\begin{equation}\label{l3}
\kappa = \frac{\zeta^2}{\eta \alpha}\,,\,\,\,
\beta=\frac{\beta_{1}}{\alpha}\,.
\end{equation}
In order to have finite solutions at the limit $y\rightarrow
\infty$, one should take the following ansatz
\begin{equation}\label{l4}
R_{\ell}(y)=y^{-\beta}\psi_{\ell}(y)\,,
\end{equation}
which transforms Eq.(\ref{l2}) into
\begin{equation}\label{l5}
\left(y\frac{d^2}{d y^2}-(2\beta-1)\frac{d}{dy}-\frac{1}{4}y +
\frac{\kappa}{2}\right)\psi_{\ell}(y)=0\,.
\end{equation}
The last equation is the same as the one that has already been
obtained in \cite{Chen1} for the  case $\ell=0$. But by
appropriate transformations this can be applied to the case
$\ell\neq 0$. Hence by following the same procedure, the solutions
of Eq.($\ref{l5}$) can be found via Laplace transforms. By
applying Laplace transform $F(s)=
\mathcal{L}(\psi)$$=\int_{0}^{\infty} e^{-sy}\psi_{\ell}(y)dy$,
\cite{Kre}, to Eq.(\ref{l5}), the following equation can be
obtained
\begin{equation}\label{la5}
\left(s^2-\frac{1}{4}\right)\frac{d}{ds}F(s)+\left[(2\beta+1)s-\frac{\kappa}{2}
\right]F(s)=0\,,
\end{equation}
which is a first order differential equation and its solutions are
in the form
\begin{equation}\label{lb5}
F(s)=N\left(s+\frac{1}{2}\right)^{-(2\beta+1)}\left(1-\frac{1}{s+\frac{1}{2}}\right)^{[\kappa-(2\beta+1)]/2}\,,
\end{equation}
where $N$ is a constant. Here
$\left(1-\frac{1}{s+\frac{1}{2}}\right)^{\kappa-(2\beta+1)}$ is a
multi-valued function. In order to have a single valued wave
function we impose the condition
\begin{equation}\label{lc5}
\kappa-(2\beta+1)=2n,\,\,\,\,n=0,1,2,3,...\,.
\end{equation}
Now considering Eq.(\ref{e10}), Eq.(\ref{l3}) and Eq.(\ref{lc5}),
the eigenvalues of the bound states can be obtained as
\begin{equation}\label{l6}
 E_{nl}=\frac{\hbar^2}{2m r_{0}^2}\left[\ell(\ell+1)C_{0}-\alpha^2\left(n+\frac{1}{2}-\frac{\zeta^2}{2\eta
 \alpha}\right)^2\right]\,,
\end{equation}
where the parameters $\zeta$, $\eta$ are given by Eq.(\ref{e10}).
This bound state energy spectrum is the same as the spectrum
obtained in \cite{BH} by using  Nikiforov-Uvarov approach. Also
for $\ell=0$ this eigenvalue relation is reduced to the relation
obtained in \cite{Chen1}.

In order to find the eigenfunctions, we should apply inverse
Laplace transforms to Eq.(\ref{lb5}). However to do so, it needs
to be expanded in power series. Following the method of
\cite{Chen1} the normalized eigenfunctions can be achieved as
\begin{equation}\label{l7}
 R_{\ell n}(y)=
 N_{n}y^{\frac{\kappa}{2}-(n+\frac{1}{2})}e^{-y/2}L_n^{\kappa-2n-1}(y)\,.
\end{equation}
$L_{n}^{\beta}$ are generalized Laguerre polynomials and $N_n$ is
given by
\begin{equation}\label{l8}
 N_{n}=\left[\frac{\alpha
 n!(\kappa-2n-1)}{\textsf{r}_0\Gamma(\kappa-n)}\right]^{1/2}.
\end{equation}

The proposed solutions are based on the Perkeris approximation and
their validity related to the magnitude of the quantum number
$\ell$. Let us now make a comparison between the calculated
eigenvalues Eq.(\ref{l6}) and the results of the two-point
quasi-rational approximation method \cite{CPM, CMP}. This method
is an extension of Pad\'{e} procedure and is reliable even for
high rotational and vibrational quantum numbers. This comparison
is made via Fig.(2) for three vibrational levels $n=0$, $n=1$ and
$n=2$ of $H_{2}$ molecule. In this figure the solid line shows the
eigenvalues obtained from Eq.(\ref{l6}) with $\alpha=1.4405$,
$\frac{\hbar^2}{2m\textsf{r}_{0}^2}=7.5416\times10^{-3} ev$ and
$D=4.7446\, ev$. The dashed lines present the eigenvalues obtained
from \cite{CMP}. It is evident from the figure that there is a
noticeable concurrence between two methods approximately for
$\ell\leq24$. By increasing the vibrational quantum number $n$,
the coincidence is elongated to smaller values of $\ell$.
Specifically the coincidence for $n=0$, $n=1$ and $n=2$ can be
seen up to $\ell=28$, $\ell=26$ and $\ell=24$, respectively.
Fig.(1) also demonstrates that the worse concurrence occurs for
$I_2$ molecule, while a better concurrence is resulted for HC1
molecule.

\begin{figure}
\centerline{\epsfig{figure=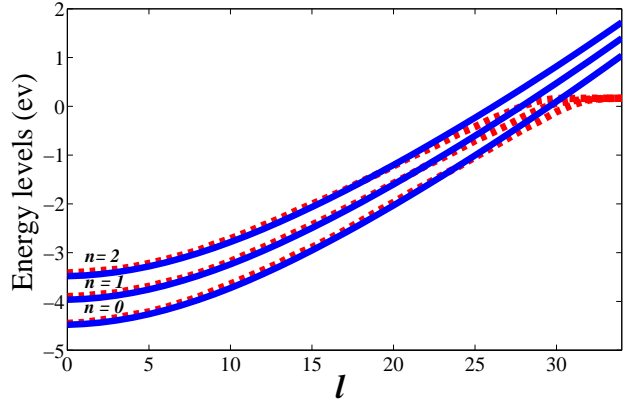,width=9.5cm}} \caption{{\small
Energy eigenvalues versus rotational quantum number $\ell$.}}
\end{figure}

%%%%%%%%%%%%%%%%%%%%%%%%%%%%%%%%%%%%%%%%%%%%%%%%%%%%%%%%%%%%%%%%%%%%%%%%%%%%
\section{Bound state solutions in $N$ dimensions}
The procedure of the preceding section for solving the schrodinger
equation with the Morse potential, can be generalized to any given
arbitrary dimension $N$. The time independent Schr\"{o}dinger
equation in $N$ dimension can be written as follows

\begin{equation}\label{d1}
\frac{-\hbar^2}{2m}\nabla_N^2\psi(\textbf{\textsf{r}})=\left[E-V(\textbf{\textsf{r}})\right]\psi(\textbf{r})
\end{equation}
where $V(\textbf{\textsf{r}})$ is an arbitrary potential. In the
presence of a spherical symmetry potential, the wave equation
Eq.(\ref{d1}) is known to be separable and can be written as
follows, \cite{N}-\cite{HHZR}

\begin{equation}\label{d2}
\psi(\textbf{r})=\textsf{r}^{-(N-1)/2}
R_{\ell}(\textsf{r})Y_{\ell_{N-2},...\ell_1}^{\ell}(\hat{\textbf{\textsf{r}}})
\end{equation}
where $Y_{\ell_{N-2},...\ell_1}^{\ell}(\hat{\textbf{\textsf{r}}})$
are generalized spherical harmonics. Here
$\hat{\textbf{\textsf{r}}}$ is the unite vector in $N$ dimension
and is supposed to be characterized with $(N-1)$ angular
coordinates. Using Eq.(\ref{d2}) in Eq.(\ref{d1}), the
Schr\"{o}dinger equation for the radial part
$R_{\ell}(\textsf{r})$ becomes

\begin{equation}\label{d3}
\frac{-\hbar^2}{2m}\left[\frac{d^2}{d
\textsf{r}^2}-\frac{\lambda^2-\frac{1}{4}}{\textsf{r}^2}\right]R_{\lambda}(\textsf{r})=[E_{\lambda}-V(\textsf{r})]R_{\lambda}(\textsf{r})
\end{equation}
where

\begin{equation}\label{d4}
 \lambda=\ell-1+\frac{N}{2}\,.
\end{equation}
Applying the Morse potential, Eq.(\ref{d3}) takes the following
form

\begin{equation}\label{d5}
\left\{\frac{d^2}{d
r^2}-\frac{\lambda^2-\frac{1}{4}}{(r+1)^2}-\frac{2m}{\hbar^2}\textsf{r}_{0}^2\left[D
(e^{-2\alpha r}-2e^{-\alpha r})-E_{\lambda}\right]\right\}
R_{\lambda}(r)=0
\end{equation}
where the variable $r$ is defined in Eq.(\ref{e5}). Using
expansion (\ref{e7}) with definitions (\ref{e8}), Eq.(\ref{d5})
can be written as

\begin{equation}\label{d6}
\left(\frac{d^2}{d r^2}- \eta_{N}^2 e^{-2\alpha r} +  2\zeta_{N}^2
e^{-\alpha r}- \beta_{1N}^2\right) R_{\lambda}(r)=0\,.
\end{equation}
Here the parameters $\eta_{N}$, $\zeta_{N}$ and $\beta_{1N}$ are
analogue to respectively $\eta$, $\zeta$ and $\beta$ defined in
three dimensional case in Eq.(\ref{e10}), and are given as follows

\begin{equation}\label{d7}
\begin{tabular}{c}
$\beta_{1N}^2= -\frac{2m \textsf{r}_{0}^2}{\hbar^2 }
E_{\lambda}+(\lambda^2-\frac{1}{4})C_{0}$
\,,\\\\
$\zeta^2_{N}  = \frac{4m \textsf{r}_{0}^2}{\hbar^2} D -\frac{1}{2}(\lambda^2-\frac{1}{4})C_{1}$\,,\\\\
$\eta^2_{N}  = \frac{2m \textsf{r}_{0}^2}{\hbar^2}
D+(\lambda^2-\frac{1}{4})C_{2}$\,.
\end{tabular}
\end{equation}
The form of Eq.(\ref{d6}) is exactly similar to Eq.(\ref{e9}) and
Hence the solutions of the former equation can be obtained via the
method of the latter one. Following the procedure of the last
section the bound state energy spectrum can be found as

\begin{equation}\label{d8}
 E_{n\lambda}=\frac{\hbar^2}{2m
 r_{0}^2}\left[\left(\lambda^2+\frac{1}{4}\right)C_{0}-\alpha^2\left(n+\frac{1}{2}-\frac{\zeta_N^2}{2\eta_N
 \alpha}\right)^2\right]\,,
\end{equation}
with
\begin{equation}\label{d9}
n=0,1,2,3,...\,.
\end{equation}
Also the corresponding normalized wave functions can be obtained
as
\begin{equation}\label{d10}
 R_{\lambda n}(y)=
 N_{n}y^{\frac{\kappa_{N}}{2}-(n+\frac{1}{2})}e^{-y/2}L_n^{\kappa_{N}-2n-1}(y)\,.
\end{equation}
Here the parameter $\kappa_{N} =\frac{\zeta_{N}^2}{\eta_{N}
\alpha} $ and $y= \frac{2\eta_{N}}{\alpha} e^{-\alpha r}$ and
$L_{n}^{\beta}$ are generalized Laguerre polynomials. Considering
Eq.(\ref{d2}) the normalization condition is given as
$\int\frac{1}{\alpha y}|R_{\lambda n}(y)|^2 dy=1 $.  Applying this
normalization condition and using the integrals of Laguerre
polynomials, the normalization constant $N_n$ can be found as
follow
\begin{equation}\label{d11}
 N_{n}=\left[\frac{\alpha
 n!(\kappa_{N}-2n-1)}{\textsf{r}_0\Gamma(\kappa_{N}-n)}\right]^{1/2}.
\end{equation}
As an example Fig.(3) shows the solutions for eigenfunctions and
eigenvalues in seven dimension.

\begin{figure}
\centerline{\begin{tabular}{cc}
\epsfig{figure=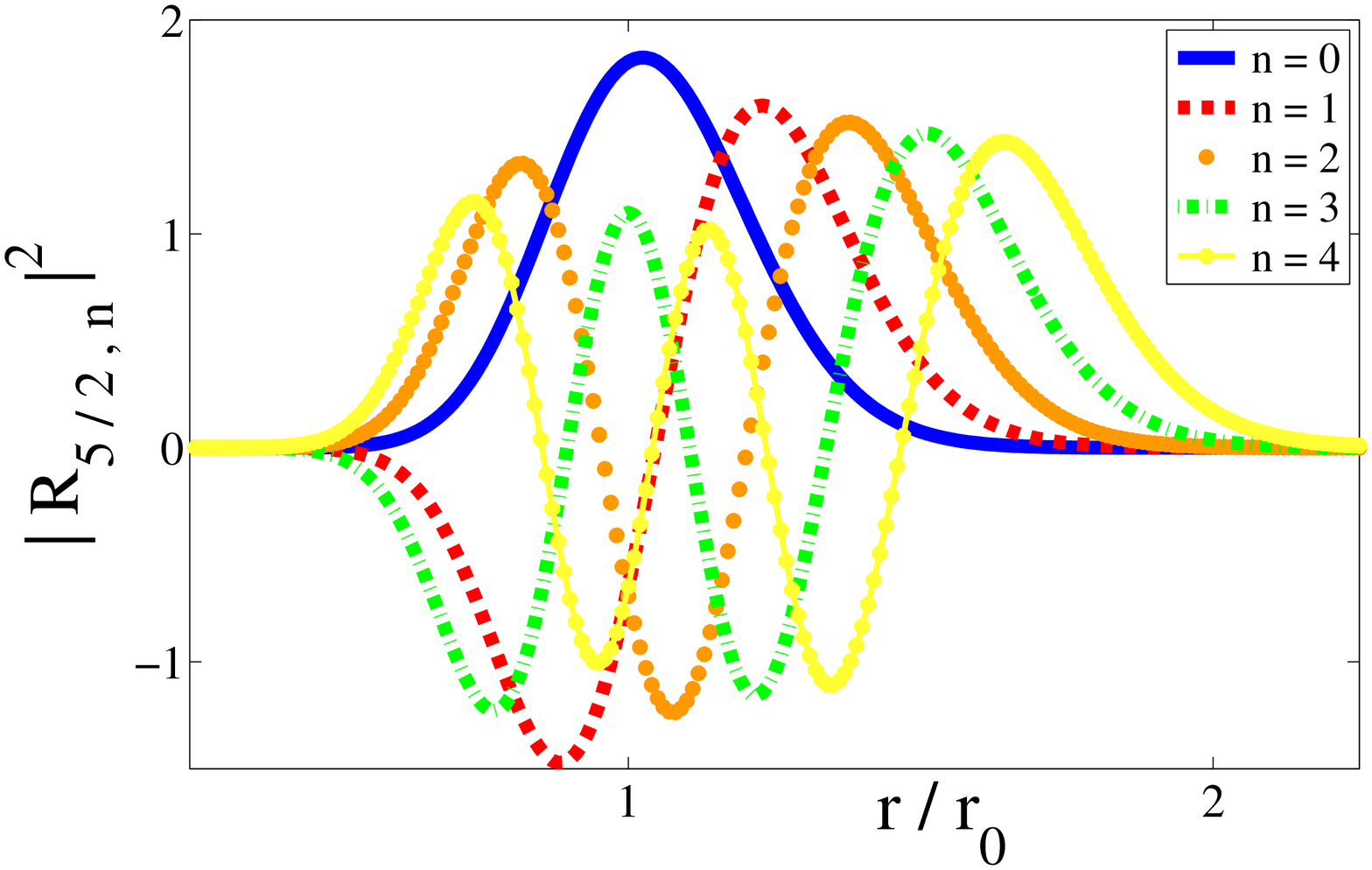,width=8.5cm}&
\epsfig{figure=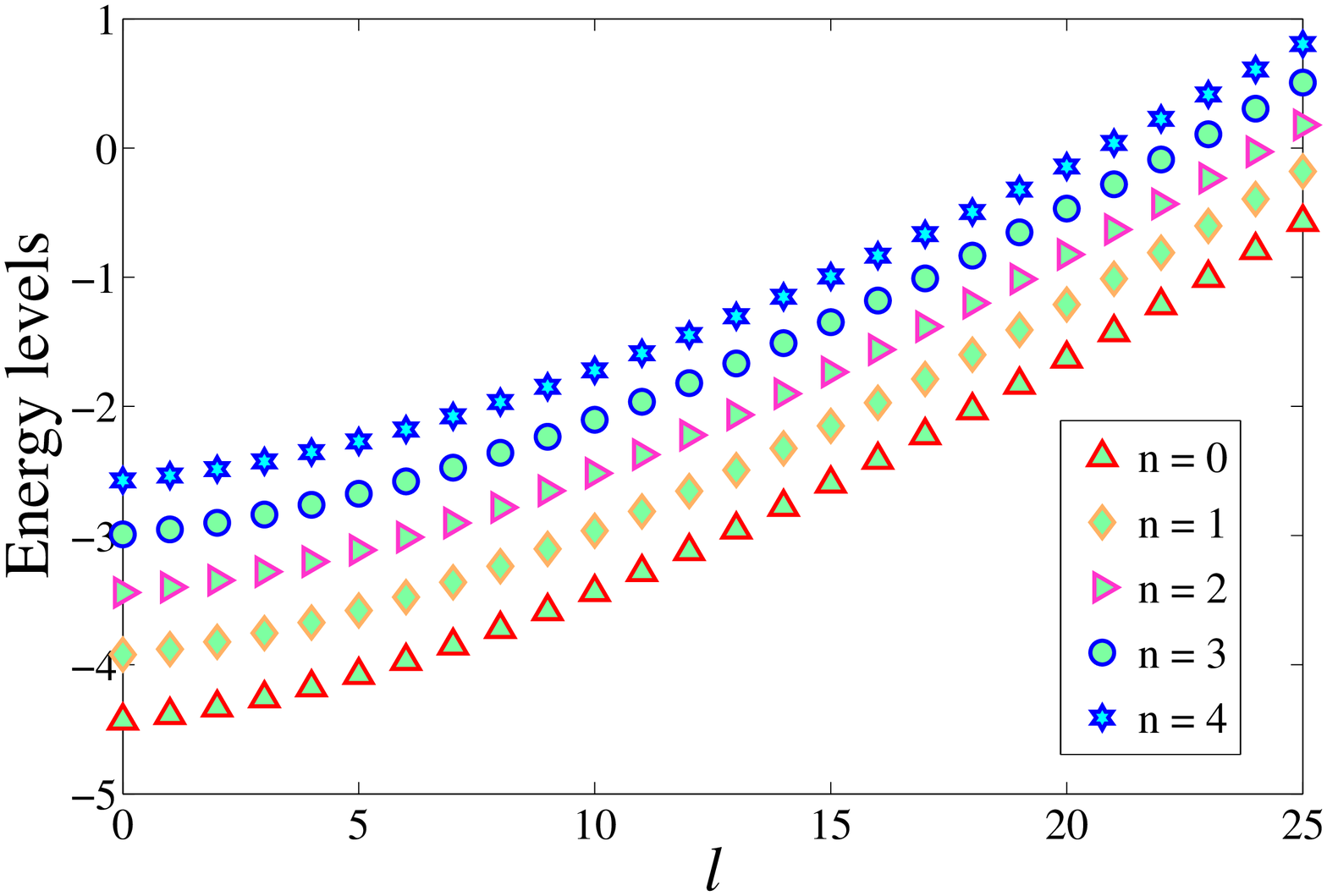,width=8.5cm}\\
\end{tabular}}
\caption{{\small The eigenvalues and eigenfunctions solutions in
seven dimensions for hydrogen atom and with $\ell=0$ and
$\lambda=5/2$.}} \label{fig1}
\end{figure}

%%%%%%%%%%%%%%%%%%%%%%%%%%%%%%%%%%%%%%%%%%%%%%%%%%%%%%%%%%%%%%%%%%%%%%%%%%%

\section{Conclusions}
This study has found an analytical solution for the energy
spectrum and wave functions of the Morse problem in arbitrary
dimension based on the Pekeris approximation and following
\cite{Chen1}, by using the Laplace transforms. The obtained
solutions in three dimensions are exactly coincides with the
results of \cite{BH} which are attained by Nikiforov-Uvarov
approach. This paper shows that the results can be justified by
comparing the calculated energy spectrum with those obtained via
the two-point quasi-rational approximation method. This comparison
in fact was undertaken to validate the Pekeris approximation which
is the basis of the proposed procedure. Nevertheless, the results
show that this approximation is not valid for high values of
quantum numbers $n$ and $\ell$ except at very low separation
distances. Despite this fact the procedure represented in this
paper is effective and succinct in its approach, and does not have
the complexities of parallel methods. Finally, this method can be
generalized in order to include higher dimensions of space.


\begin{thebibliography}{99}

\bibitem{CY}         C. D. Yang, Chaos Soliton. Fract., 37 (2008)
                     962.

\bibitem{HW}          H. Haken and H. C. Wolf, Molecular Physics and Elements
                      of Quantum Chemistry: Introduction to Experiments and
                      Theory, Springer, Berlin, 1995.

\bibitem{FM}          C. Frankenberg, J. F. Meiring, M. van Weele, U.
                      Platt and T. Wagner, Science 308
                      (2005) 1010.

%%%%%%%%%%%%%%%%%%%%%% l=0 %%%%%%%%%%%%%%%%%%%%%%%%%%%%%%%%%%%%%%%%%%%%%%%%%
\bibitem{Morse}       P. M. Morse, Phys. Rev. 34 (1929)
                       57.

\bibitem{Flugge}      S. Fl\"{u}gge, Practical Quantum Mechanics,
                      Springer, Berlin, 1974.

\bibitem{IS}          S. M. Ikhdair and R. Sever, Appl. Math. Comput. 218 (2012)
                      10082.

\bibitem{Pe}          C. L. Pekeris, Phys. Rev. 45 (1934) 98.

%%%%%%%%%%%%%%%%%%%%%%%%%%%%%%%%%%%%%% factorization method %%%%%%%%%%%%%%%%

\bibitem{IH}          L. Infeld and T. E. Hull, Rev. Mod. Phys. 23 (1951) 21.

\bibitem{Ba}          A. B. Balentekin,  Phys. Rev. A 57(1998) 4188.

\bibitem{DLF}         S. H. Dong, R. Lemus and A. Frank, Int. J. Quantum Chem. 86 (2002)
                      433.
%%%%%%%%%%%%%%%%%%%%%%%%%%%%%%%%%%%%%% path integral formulation%%%%%%%%%%%%

\bibitem{G1}          C. Grosche, J. Phys. A-Math. Gen. 28(1995)
                      5889.

\bibitem{G2}          C. Grosche, J. Phys. A-Math. Gen. 29
                      (1996) 365.

\bibitem{KS}          N. Kandirmaz and R. Sever, Chin. J. Phys. 47(1)(2009) 46.

%%%%%%%%%%%%%%%%%%%%%%%%%%%%%% super symmetric %%%%%%%%%%%%%%%%%%%%%%%%%%%%%%

\bibitem{CKS}         F. Cooper, A. Khare and U. Sukhatme, Phys. Rep. 251 (1995) 267.

\bibitem{BM}          M. G. Benedict and B. Moln\'{a}r, Phys. Rev. A 60 (1999) 1737.

\bibitem{FR}          E. D. Filho and R. M. Ricotta, Phys. Lett. A 269 (2000)
                      269.

\bibitem{RD}          F. R. Silva and E. D. Filho, J. Chem. Phys. Lett.
                      498 (2010) 198.

\bibitem{Gran}        Y. Grandati,  Phys. Lett. A 376 (2012)
                      2866.

%%%%%%%%%%%%%%%%%%%%%%%%%%% algebraic way %%%%%%%%%%%%%%%%%%%%%%%%%%%%%%%%%%%


\bibitem{BGQ}        B. Bagchi, P. S. Gorain and C. Quesne, Mod. Phys. Lett. A 21 (2006) 2703.

\bibitem{FD}         G. F. Wei and S. H. Dong, Europhys. Lett. 87 (4) (2009) 40004.

\bibitem{YHB}        S. A. Yahiaoui, S. Hattou and M. Bentaiba, Ann. Phys.  322 (2007) 2733

\bibitem{Co}         I. L. Cooper, J. Phys. A-Math Gen. 26 (1993)
                     1601.

\bibitem{CG}         I. L. Cooper and R. K. Gupta, Phys. Rev. A 52 (1995)
                     941.

\bibitem{GDD}        Z. H. Geng, Y. Dai and S. L. Ding, J. Chem. Phys. 278 (2002)
                     119.

%%%%%%%%%%%%%%%%%%%%%%%%%% power series %%%%%%%%%%%%%%%%%%%%%%%%%%%%%%%%%%%%%%

\bibitem{JDS}       J. Yu, S. H. Dong and G. H. Sun, Phys. Lett. A 322 (2004)
                    290.

\bibitem{DS}        S. H. Dong and G. H.  Sun, Phys. Lett. A 314 (2003)
                    261.

\bibitem{Z}         A. J. Zakrzewski, Comput. Phys. Commun. 175 (2006)
                    397.


%%%%%%%%%%%%%%%%%%% two-point quasi-rational approximation method %%%%%%%%%%%%%

\bibitem{CPM}       E. Castro, J. L. Paz and P. Martin, J. Mol. Struc-Theochem  769 (2006) 15.

\bibitem{CMP}       E. Castro, P. Martin and J. L. Paz, Phys. Lett. A 364 (2007)
                    135.

%%%%%% 1/N %%%%%%%%%%%%%%%%%%%%%%%%%%%%%%%%%%%%%%%%%%%%%%%%%%%%%%%%%%%%%%%%%%%%%

\bibitem{Lai}       C. H. Lai, J. Math. Phys. 28 (1987)
                    1801.

\bibitem{Atag}      S.  Atag, Phys. Rev. A 37 (1988) 2280.

\bibitem{Imbo}      T. D. Imbo and U. P. Sukhatme, Phys. Rev. Lett. 54 (1985)
                    2184.

\bibitem{Bag}       M. Bag, M. M. Panja and R. Dutt, Phys. Rev. A 46 (1992) 6059.

%%%%%%%%%%%%%%%%%%%%%%%%%%%%%%% matrix %%%%%%%%%%%%%%%%%%%%%%%%%%%%%%%%%%%%%%%%%%

\bibitem{OCS}       Y. Ou, Z. Cao and Q. Shen, Phys. Lett. A 318 (2003)
                    36.
\bibitem{S}         H. Sun, Phys. Lett. A 338 (2005)
                    309.
\bibitem{NA}        I. Nasser,  M. S. Abdelmonem , H. Bahlouli and  A.  D. Alhaidari,
                    J. Phys. B-At. Mol. Opt. 40 (2007) 4245.

%%%%%%%%%%%%%%%%%%%asymptotic iteration method%%%%%%%%%%%%%%%%%%%%%%%%%%%%%%%%%%%

\bibitem{CHS1}      H. Ciftci, R. L. Hall and N. Saad, J. Phys. A-Math. Gen. 36 (2003)
                    11807.

\bibitem{CHS2}      H. Ciftci, R. L. Hall and N. Saad, J. Phys. A-Math. Gen. 38 (2005)
                    1147.

\bibitem{BB}        O. Bayrak and I. Boztosun, J. Mol. Struc-Theochem 802 (2007) 17.


%%%%%%% Uvarov %%%%%%%%%%%%%%%%%%%%%%%%%%%%%%%%%%%%%%%%%%%%%%%%%%%%%%%%%%%%%%%%%%

\bibitem{BH}        C. Berkdemir and J. Han, J. Chem. Phys. Lett. 409 (2005)
                    203.

\bibitem{SI}        S. M. Ikhdair, J. Chem. Phys. 361 (2009) 9.

%%%%%%%%%%% Laplace transformation %%%%%%%%%%%%%%%%%%%%%%%%%%%%%%%%%%%%%%%%%%%%%%

\bibitem{Kre}       E. Kreyszing, Advanced Engineering Mathematics, John Wiley and
                    Sons, 1979.

\bibitem{Sch}       E. Schr\"{o}dinger, Ann. Phys. 384 (1926)
                    361.

\bibitem{En1}       M. J. Englefield, J. Austral. Math. Soc. 8
                    (1968) 557.

\bibitem{En2}       M. J. Englefield, J. Math. Anal. and Appl. 48 (1974)
                    270.

\bibitem{SD}        R. A. Swainson  and G. W. F. Drake,  J. Phys. A-Math. Gen.
                    24 (1991) 79.

\bibitem{Chen1}     G. Chen, phys. Lett. A 326 (2004) 55.

\bibitem{Chen3}     G. Chen, Chin. Phys. 14 (6) (2005) 1075.

\bibitem{AS}        A. Arda and R. Sever, J. Math. Chem. 50 (2012)
                     971.

\bibitem{PC}        D. R. M. Pimentel and A. S. de Castro, Eur. J. Phys. 34 (2013)
                    199.
%%%%%%%%%%%%%%%%%%%%%%%%%%%%%%%%%%%%%%%%%%%%%%%%%%%%%%%%%%%%%%%%%%%%%%%%%%%%%%%%%%%

\bibitem{N}         M. M. Nieto, Phys. Lett. A 293 (2002)
                    10.

\bibitem{Chen2}     G. Chen, Phys. Lett. A 329 (2004)
                    22.

\bibitem{CC}        G. Chen and Z. D. Chen, Phys. Lett. A 331 (2004)
                    312.

\bibitem{HHZR}      H. Hassanabadi, M. Hamzavi, S. Zarrinkamar and A. A.
                    Rajabi, Int. J. Phy. Sci. 6 (3)
                    (2011) 583.

\end{thebibliography}
\end{document}